\documentclass[conference,letterpaper]{IEEEtran}

\usepackage{cite}
\usepackage{algorithmic}
\usepackage{graphicx}
\usepackage{textcomp}
\usepackage{xcolor}
\def\BibTex{{\rm B\kern-.05em{\sc i\kern-.025em b}\kern-.08em
    T\kern-.1667em\lower.7ex\hbox{E}\kern-.125emx}}
\usepackage[utf8]{inputenc} 
\usepackage[T1]{fontenc}
\usepackage[english]{babel}
\usepackage{csquotes}
\usepackage{mleftright}
\usepackage[cmex10]{amsmath}  
\usepackage{amsfonts}
\usepackage{amssymb}
\usepackage{blkarray}            
\usepackage[caption=false,font=footnotesize]{subfig}                           
\usepackage{stfloats}
\usepackage{tikz}

\usepackage{tabularx} 
\usepackage{multirow}

\usepackage{array}
\usepackage{kantlipsum}

\newtheorem{theorem}{\textbf{Theorem}}[section]
\newtheorem{proposition}[theorem]{\textbf{Proposition} }
\newtheorem{lemma}[theorem]{\textbf{Lemma}}

\newtheorem{corollary}[theorem]{ \textbf{Corollary}}

\newcommand{\F}{\mathbb{F}}
\newcommand{\G}{\mathbb{G}}
\newcommand{\LL}{\mathbb{L}}
\newcommand{\N}{\mathbb{N}}
\newcommand{\R}{\mathbb{R}}
\newcommand{\Z}{\mathbb{Z}}

\newcommand{\bA}{\mathbf{A}}
\newcommand{\bB}{\mathbf{B}}
\newcommand{\bC}{\mathbf{C}}
\newcommand{\bH}{\mathbf{H}}

\newcommand{\bP}{\mathbf{P}}

\newcommand{\bu}{\mathbf{u}}
\newcommand{\bv}{\mathbf{v}}

\newcommand{\wt}{\mathsf{wt}}
\newcommand{\qed}{\hfill $\square$}

\newcommand{\calC}{\mathcal{C}}
\begin{document}

\title{On the Generalization of Kitaev Codes as Generalized Bicycle Codes} 

\author{
\IEEEauthorblockN{François Arnault\IEEEauthorrefmark{1}, Philippe Gaborit\IEEEauthorrefmark{1}, Nicolas Saussay\IEEEauthorrefmark{1}}
\IEEEauthorblockA{\IEEEauthorrefmark{1} XLIM, UMR 7252, Université de Limoges\\ 123, Avenue Albert Thomas, 87000 Limoges, France\\ \{arnault, philippe.gaborit, nicolas.saussay\}@unilim.fr}
}

\maketitle

\begin{abstract}
\nocite{Shor95, NC2010, K03,BD07, LZ22, PK22, IBM24, HHO21, CS96, S96, KP13, MMM04, PW22}
Surface codes have historically been the dominant choice for quantum error correction due to their superior error threshold performance. However, recently, a new class of Generalized Bicycle (GB) codes, constructed from binary circulant matrices with three non-zero elements per row, achieved comparable performance with fewer physical qubits and higher encoding efficiency.

In this article, we focus on a subclass of GB codes, which are constructed from pairs of binary circulant matrices with two non-zero elements per row.
 We introduce a family of codes that generalizes both standard and optimized Kitaev codes. 
These codes exhibit parameters of the form $ [| 2n , 2, \geq \sqrt{n} |] $ where $ n$ is a factor of $ 1 + d^2 $.  
For code lengths below 200,  our analysis yields $21$ codes, including $7$ codes from Pryadko and Wang's database, and unveils $14$ new codes with enhanced minimum distance compared to standard Kitaev codes. Among these, $3$ surpass all previously known weight-4 GB codes for distances $4$, $8$, and $12$.

\end{abstract}

\section{Introduction}

\subsection{Motivation}
Quantum computers hold the promise of revolutionizing computation by efficiently solving problems that are intractable for standard computers, as demonstrated Shor's algorithm \cite{Shor95}. However, the inherent fragility of quantum information (decoherence) and the presence of environmental noise make quantum computers susceptible to errors during computation. These errors accumulate over time, potentially leading to incorrect syndrome measurements and erroneous results.

To enable long-duration quantum computation, quantum error correction is essential. 
This process involves employing quantum codes, syndrome measurement circuits, and decoding algorithms to detect and correct errors as they occur. As errors can accumulate even during syndrome measurement and decoding, the error correction process must be both fast and reliable. To ensure reliable quantum computation, the physical error rate must be kept below a critical threshold, allowing for effective error suppression \cite{NC2010}.

\subsection{Previous results:}
Kitaev codes, introduced in 2003, were the first quantum low-density parity-check (qLDPC) codes with minimum distance scaling linearly with the square root of the code length \cite{K03}. Since then, various generalizations of Kitaev codes have been developed in an effort to improve their parameters. Notable examples include Kitaev's optimized 45-degree rotated codes \cite{BD07}, which offer enhanced minimum distance, and surface codes, which provide higher rates.
However, neither of these generalizations has managed to surpass the square root scaling of the minimum distance.

Until recently, all known qLDPC codes were limited by a square root scaling of the minimum distance. However, in recent years, significant progress has been made, with the theoretical existence of asymptotically good qLDPC codes, possessing positive rate and linear minimum distance scaling, being established \cite{LZ22, PK22}.
While these codes offer superior theoretical performance, their practical implementation remains challenging compared to surface codes, which exhibit high error thresholds and efficient decoding algorithms.

\subsection{Contributions}

Currently, Generalized Bicycle (GB) codes, denoted as$(a,b)$-GB when constructed from circulant matrix pairs with $a$ and $b$ non-zero elements per row, offer the most promising alternative to surface codes for practical implementation.

 In their 2024 paper \cite{IBM24}, the IBM research team constructed a family of $(3,3)$-GB codes, from pairs of circulant matrices with three non-zero elements per row, achieving comparable error thresholds with significantly lower encoding overhead.
 

 $(2,2)$-GB codes, which generalize Kitaev codes and their optimized versions \cite{KP13}, offer the advantages of simple structures, planar layouts, and local weight-four generators, making them highly attractive for physical implementation. 

Despite, the good $(2,2)$-GB codes identified in the numerical searches for lengths $2p$ where  $ p$ is a prime for which $2$ is a primitive root of unity \cite{PW22}, the literature lacks $(2,2)$-GB codes with minimum distance scaling linearly with the square root of the code length, beyond Kitaev codes. 

Building on Haah, Hastings, and O'Donnell's suggestion \cite{HHO21} that toric code geometry alterations could yield length $N$ codes with minimum distance scaling as  $ \sqrt{cN}$ where $c > \frac{1}{2}$ we introduce a new class of $(2,2)$-GB codes based on $2D$ lattices with acute angles between basis vectors, achieving a minimum distance intermediate between standard and optimized Kitaev codes.

While Kitaev-optimized codes are restricted to parameters of the form $[| 1 + d^2, 2, d |]$ with $d$ odd, our proposed codes offer a more flexible parametrization of the form $[| 2n, \, 2, \, \geq \sqrt{n} |] $ where $n$ can be any divisor of $ 1 + d^2$ for any positive $d$.

Notably, for code lengths below 200, we identify 21 codes, encompassing 7 from Pryadko and Wang's database \cite{PW22}, and reveal 14 novel codes that outperform standard Kitaev codes in minimum distance. Among these, 3 surpass all previously known $(2,2)$-GB codes for distances 4, 8, and 12. This provides more options for designing small-scale quantum error correction codes, which is currently a critical requirement for the practical implementation of quantum computing.

\subsection{High-level overview}

Our approach to constructing $(2,2)$-GB codes that outperform Kitaev codes involves four key steps:

\begin{itemize}

\item \textbf{Lattice Association:} Mapping $(2,2)$-GB codes of length $ 2n$ to 2D lattices $\LL$ by exploiting the spatial arrangement of non-zero elements in the defining matrices. (Sections \ref{sec: background on lattices} and \ref{section:GB_and_graph_theory})

\item \textbf{Graph Representation:} Representing qubits of the $(2,2)$-GB code on the surface of the 2D torus $\Z^2/\LL$. (Section \ref{section:GB_and_graph_theory})

    \item \textbf{Minimum distance bound:} Using the graph formalism, to lower bound the minimum distance of the code by the shortest non-zero lattice's  vector length.
    (Sections \ref{sec: Reinterpretation of the minimum distance}, \ref{sec: from G-walks to Z^2-walks}, \ref{sec: from Z^2 walks to G walks} and \ref{sec: end of main theorem}) 
    
    \item \textbf{Arithmetic Analysis:} Leveraging arithmetic properties to identify specific cases where the lattice's shortest non-zero vector length exceeds $ \sqrt{n}$, the square root of the code length divided by two. (Section \ref{section: application of the main theorem})
\end{itemize}

\subsection{Organization of the paper }
The paper is divided into three sections. Section~\ref{sec: notations and relevant facts} provides the needed background information on GB codes. Section~\ref{section: Proof of the main theorem} presents the proof of the aforementioned bound on the minimum distance. Finally, Section~\ref{section: application of the main theorem} explains how to explicitly construct the $(2,2)$-GB codes that outperform Kitaev codes.

\section{Notations and relevant facts}
\label{sec: notations and relevant facts}
\subsection{Background on quantum codes}

In this section, we start by reviewing CSS codes and then we introduce the notions of circulant matrices and GB codes.
\vspace{0.1cm}

\subsubsection{CSS codes}

Calderbank-Shor-Steane (CSS) codes \cite{CS96, S96} form an important class of quantum error-correcting codes. These codes can be described by two binary matrices $\bH_X$ and $\bH_Z$ whose row spaces are orthogonal, i.e., \mbox{$\bH_X \bH_Z^\intercal = \mathbf{0}$}. The code length $n$, representing the number of physical qubits, is the number of columns of $\bH_X$ (or of $\bH_Z$). The code dimension $k$, corresponding to the number of logical qubits, is given by $k = n - \mathrm{rank} \, \bH_X - \mathrm{rank} \, \bH_Z$. The minimum distance $d$ of the code is defined as the minimum of $d_X$ and $ d_Z$ where $d_X$ (resp. $d_Z$) denotes the smallest Hamming weight of a codeword of $\ker \bH_X$ (resp. $\ker \bH_Z$) that is not in $ rs(\bH_Z) $ the row space of $\bH_Z$ (resp. not in $rs(\bH_X$) the row space of $ \bH_X$).
\vspace{0.1cm}

\subsubsection{Circulant matrices}

Let $n>1$ be an integer and let $A = \sum_{i=0}^{n-1} a_i x^i \in \F_2[x]_{\leq n - 1} $ be a polynomial. The circulant matrix associated to $A$ and $n$ is given by
\begin{equation*}
Circ(A, \, n)  = \begin{pmatrix} 
a_0     & a_{n-1} & \dots  & a_{1}  \\
a_{1}   & a_0     & \ddots & \vdots \\
\vdots  & \ddots  & \ddots & a_{n - 1} \\
a_{n-1} & \dots   & a_{1}  & a_0
\end{pmatrix}.
\end{equation*}
Since any circulant matrix is a polynomial in the cyclic permutation matrix given by the companion matrix of $x$, the product of two circulant matrices is commutative. 
\vspace{0.1cm}

\subsubsection{Generalized bicycle (GB) codes}
\vspace{0.1cm}

The generalized bicycle (GB) codes introduced by Pryadko and Kovalev in their 2013 paper \cite{KP13} extend  Mackay et al.'s bicycle codes \cite{MMM04}.

While bicycle codes' parity check matrices  $\bH_X = \bH_Z = [\bA, \bA^\intercal ]$  are defined using a single circulant matrix $\bA$, GB codes employ two circulant matrices $\bA$ and $\bB$ to construct their parity-check matrices. 
Specifically, for binary polynomials $A, \, B \in \F_2[x]_{\leq n - 1 }$, the GB code $GB(A,B,n)$ is defined by the parity-check matrices $\bH_X = [\bA, \bB]$ and $\bH_Z = [\bB^{\intercal}, \bA^{\intercal}] $ where $ \bA = Circ(A, \, n)$ and $ \bB = Circ(B, \, n)$. Due to the equivalence of $GB(A,B,n)$ and $GB(B,A,n)$, the code parameters are \mbox{$ [|2n, \, 2 \deg (\gcd(A, B, x^n - 1)), \, d_X = d_Z|]$}.
\vspace{0.1cm}

In this work, we aim to maximize the minimum distance of $(2,2)$-GB codes generated by two binary polynomials $A$ and $ B$ with exactly two non-zero terms. 
\vspace{0.1cm}

As sums up the next proposition, in our study of large distance $(2,2)$-GB codes, we only need to investigate cases where $ A, B \in \F_2[X]_{\leq n - 1}$ have non-zero constant terms.
\vspace{0.1cm}

\begin{proposition}  
\label{prop:equivalent_GB_codes}
Let $ n > 0 $ and $ A, B \in \F_2 \left[x\right]_{\leq n - 1}$ be two polynomials. Let $ i, j , k \in \N $ be three integers. 
\vspace{0.05cm}
\begin{itemize}
    \item If$ k$ is coprime with $n$ then $ GB(A,B,n)$ and \mbox{$GB( A(x^k) \mod  x^n - 1, B(x^k) \mod x^n - 1, n)$} are equivalent \cite{PW22}. In particular, when  $r \geq 1 $ and $ n$ are coprime  we have:
 \begin{equation*}
      d(GB(1 + x^r, 1 + x^{s}, \,  n)) = d(GB(1 + x, 1 + x^\alpha, \, n))
 \end{equation*}
 where $ \alpha =  \, \, st \, \, mod \, n  $ with $ t = ( r \, \, mod \, n)^{-1} $ 
\vspace{0.15cm}

    \item $GB(Ax^i \mod x^n - 1, Bx^j \mod x^n - 1,n )$  and $GB(A,B,n)$ are equivalent 
\end{itemize}

\end{proposition}

\vspace{0.1cm}

\textbf{Proof of the $ 2^{nd} $ point:}
Let $ R_i $ and $S_j$ be the residue of $Ax^i$ and $ Bx^j$ modulo $ x^n - 1$. We set 
$ \bA = Circ(A,n), \, \bB = Circ(B,n)$ and $ \bC =  Circ(x,n)$.
\vspace{0.1cm}

The map $ h$ defined by

\begin{equation*}
   \begin{array}{ccccl}
h & : & \F_2^{2n} & \to & \F_2^{2n}\\
 & & \begin{pmatrix}
     u \\ v
 \end{pmatrix}  & \mapsto & \begin{pmatrix}
     \bC^iu \\ \bC^j v
 \end{pmatrix} 
\end{array} 
\end{equation*}
\vspace{0.1cm}

 induces  isometries (for the Hamming weight) between:
\vspace{0.05cm}

\begin{itemize}
    \item the right kernels of $  \left[ \bA  , \, \bB \right] $ and $  \left[ \bC^i \bA, \,  \bC^j \bB \right] $ 
\vspace{0.05cm}
    \item the row spaces of $  \left[ \bB^\intercal , \,  \bA^\intercal \right] $ and the row space of $ \left[ (\bC^j \bB)^\intercal , \,  (\bC^i \bA)^\intercal \right]$

\end{itemize}
\vspace{0.1cm}

Since $ Circ(R_i, n )= \bC^i \bA$ and $ Circ(S_j, n ) = \bC^j \bB$ then the two codes $ GB(A,B,n)$ and $GB(R_i, S_j, n)$ have the same parameters.
\qed
\vspace{0.1cm}

\subsection{Background on lattices}
\label{sec: background on lattices}

This section presents the theoretical foundation for our lower bound on the minimum distance of $(2,2)$-GB code.

\vspace{0.1cm}

\subsubsection{Lattices} A 2D-lattice is a set
$\LL = \Z e_1 \oplus \Z e_2 \subseteq \R^2 $ where $ (e_1, \, e_2) $ is a $\R$-basis of $ \R^2$. Its determinant $ det(\LL) $ given by $  |det(e_1, \, e_2)| $ only depends on $\LL$ and not on the basis $(e_1, e_2)$. When $ \LL $ is a 2D sublattice of $\Z^2 $, its determinant is equal to the cardinality of the quotient $ \Z^2 / \LL $.
\vspace{0.1cm}

To any $(2,2)$-GB  generated by two polynomials \mbox{$ A = 1 + x^u$,  $B = 1 + x^v \in \F_2 \left[x \right]_{ \leq n-1}$} one can associate a lattice  using $u$ and $ v$.
\vspace{0.1cm}

\subsubsection{Lattice associated to a $(2,2)$-GB code}
\vspace{0.1cm}

 The 2D-lattice associated to $GB(A,B,n)$ is given by the Kernel of the $ \Z-$linear application  
\begin{equation*}
   \begin{array}{ccccl}
\Phi  & : & \Z^2 & \to & \Z / n\Z\\
 & & (x, y)  & \mapsto & xu  + yv \mod  n  
\end{array} 
\label{eq:Phi}
\end{equation*}

which is equal to $ \Z(n, 0) \oplus \Z( - \alpha, 1)$ when $ A = 1 + x $ and $ B = 1 + x^\alpha $.
\vspace{0.1cm}

Our main result states that for any vector $\bv$ in the kernel of $[\bA, \bB]$ that is not a linear combination of the rows of $[\bB^{\intercal}, \bA^{\intercal}]$, the Hamming weight of $\bv$ is bounded below by the length of the shortest non-zero vector in the kernel of $ \Phi$.
\vspace{0.1cm}

\begin{theorem} 
For two integers $ n \geq 6 $ and $ 1  \leq \alpha  \leq n -1$  the minimum distance $ d_{min} $ of $ GB(1 + x, \, 1 + x^\alpha,\, n)$ verifies:
    \begin{equation}
        d_{min} \geq \lambda ( \LL) = \min \{ \, ||\bu|| \, \, \vert \, \, \bu \in \LL \backslash \{ 0 \} \} 
        \label{borne_inf_distance_GB}
    \end{equation}
where $ \LL =\Z(n, 0) \bigoplus \Z( - \alpha, 1) $ and $ || \cdot || $ is the Euclidean norm.
\label{thm:main_theorem}
\end{theorem}
\vspace{0.1cm}

Our proof strategy for this theorem is the following : 

\begin{itemize}
    \item \textbf{Graph reinterpretation:} We construct a graph on the vertices of $ \LL$ in a way that each vector $ \bv $ of $ \ker [\bA, \bB]$  corresponds to a graph cycle $ \tilde{v} $ whose length is $  \wt(\bv) $ the Hamming weight of $\bv $. 

    \item \textbf{Lower bound on the cycle length:} We prove that when $ \bv $ does not belong to the row-space of $[\bB^{\intercal}, \bA^{\intercal}]$ then the length of $ \tilde{v}$ is lower bounded by the $ \lambda(\LL) $ the shortest length of a non-zero vector of the lattice associated to $GB(A, B, n)$.

\end{itemize}

\section{Proof of Theorem~\ref{thm:main_theorem}}
\label{section: Proof of the main theorem}
Throughout this section, we adopt the following notations:
\begin{itemize}
    \item Let $ n \geq 6$ and $ \alpha \in [| 1, \, n - 1 |] $ be two integers

    \item  Let $ A = 1 + x $, $ \, \bA = Circ(A,n)$ 

    \item  Let $ B = 1 + x^\alpha$, $\, \bB  = Circ(B,n) $ 

    \item Let $\LL = \Z(n,0) \bigoplus \Z(-\alpha,1)  $

\end{itemize}
\vspace{0.1cm}

To prove the main theorem in the non-trivial cases, where $ 1 < \alpha < n  - 1$, we will use a graph-theoretic perspective. First, we build a graph on the vertices of $ \Z^2 / \LL \cong \Z/n \Z $.

\subsection{Graph construction}
\label{section:GB_and_graph_theory}

We reinterpret the matrix $\left[ \bA , \, \bB \right]$ which has two non-zero elements per column  as the vertex-edge incidence matrix of the  undirected graph $\G$ whose vertices and edges are given by: 
\vspace{0.1cm}

\begin{itemize}
    \item \textbf{Vertices} = $ \{ x \, | \, x \, \in \Z / n\Z \}$

    \item \textbf{ Edges}  = $ \{ h_i \, \vert \, \, 0 \leq i < 2n \}$ such that for $ 0 \leq k < n $:

    \begin{itemize}
        \item $ h_k = \{ k \, mod \, n, \, k + 1 \, mod \, n \}$

        \item $ h_{n +k } = \{ k  \, mod \, n, \, k + \alpha \, mod \, n \} $
   
    \end{itemize}
\end{itemize}

Thus, any vector $\bv = (v_0, \dots, v_{2n - 1}) \in \F_2^{2n} $ corresponds to
the set of edges $   e_v = \{ h_i \, \vert \, 0 \leq i < 2n, \, v_i = 1 \}$. 
Since $ 1 < \alpha < n - 1  $, the Hamming weight of $ \bv$ is equal to the cardinality of $ e_v $. Although, the sum of vectors $ v_1 +  v_2$ technically corresponds to the symmetric difference between $ e_{v_1}$ and $ e_{v_2}$, in the following sections we will simply refer to their symmetric difference as their "sum".
\vspace{0.1cm}

Now let's give a graph-theoretic interpretation of the minimum distance of $ GB(A,B,n )$. 

\subsection{Reinterpretation of the minimum distance}
\label{sec: Reinterpretation of the minimum distance}
By associating the rows and kernels of the parity-check matrices $\bH_X$ and $\bH_Z$ with edge subsets of $ \G $, we show that the minimum distance is equivalent to the length of a cycle in this graph.
\vspace{0.1cm}

\textbf{Reinterpretation of $ rs(\bH_X)$ and $ \ker \bH_X$:}

\begin{itemize}
    \item For $ p \in \Z / n\Z $ \textbf{cocycle($p$)}, the $ p^{\text{th}}$ row of $\bH_X$  corresponds to:
\begin{equation*}
    \{     p  \leftrightarrow (p + 1) , \, \,    p  \leftrightarrow (p - 1) , \, \,   p  \leftrightarrow (p + \alpha ),\,  \,  p  \leftrightarrow (p -  \alpha)  \}
    \label{eq:cocycle(x)}
\end{equation*}

\item  The cycles, the elements of $ \ker \bH_X$, are orthogonal to all cocycles. Moreover, each cycle $\calC$ is characterized by the property that each vertex $ v \in \Z / n\Z$ belongs to an even number of edges of $\calC$.
\end{itemize}
\vspace{0.1cm}

\textbf{Reinterpretation of $ rs(\bH_Z)$ and $ \ker \bH_Z$:}

\begin{itemize}
    \item For $ p \in \Z / n\Z $, \textbf{face($p$)},  the $ p^{\text{th}}$ row of $\bH_Z $  corresponds to:
\begin{equation*}
    \{ \, p \leftrightarrow (p + 1) \leftrightarrow( p +  1 + \alpha) \leftrightarrow (p + \alpha) \leftrightarrow p \, \} 
    \label{eq:face(x)}
\end{equation*}
\end{itemize}

Thus, the minimum distance of $GB(A, B, n)$ can be thought of as the shortest length of a cycle of $ \G$ that is not a sum of faces. Additionally, we notice that such a cycle must be connected and must not contain sub-cycles.
\vspace{0.1cm}

We decompose the proof of Theorem \ref{thm:main_theorem} on the minimum distance lower bound into two steps:
\begin{itemize}
\item To any cycle $ \calC$ of $ \G $, we associate a lattice element $ \bP \in \LL $ with a Euclidean norm smaller than the length of $ \calC$.

\item We demonstrate that if $\calC$ is not sum faces, then $ \bP \neq (0,0) $
\end{itemize}

In the following section, we define a mapping from the set of walks in $\G$ to $ \Z^2$, that maps cycles of $\G$ to elements of $\LL$.
\subsection{ From $ \G$-walks to $ Z^2$-walks }
\label{sec: from G-walks to Z^2-walks}

A walk $ \calC$ of length $ r \geq 1$ in $ \G $ is a sequence of edges  $ e_1, \dots, \, e_r $ such that consecutive edges share a common vertex. When $ e_i = \{ C_i, \, C_{i +1} \} $ for $ 1 \leq i \leq r $, we can specify the orientation of $ \calC$ by writing it as a sequence of vertices:
\begin{equation*}
 \calC \, : \, C_0 \xrightarrow{} \dots \xrightarrow{}  C_{r}
\end{equation*}

Since $ 1 < \alpha < n - 1 $ and $ n \geq 6 $, any walk of $ \G$ could have four different types of edges:
\begin{itemize}
\item Either $ x \xrightarrow[]{} x + 1 $ or $ x + 1 \xrightarrow{} x$

\item Either $ x \xrightarrow{ } x + \alpha $ or $ x + \alpha \xrightarrow{} x $
\end{itemize}
\vspace{0.1cm}

From any given $\G$-walk $ \calC$, we can construct a walk $\gamma_\calC \, : \, P_0 \xrightarrow{} P_1 \xrightarrow{} \dots \xrightarrow{} P_r \, $ in $\Z^2$ that mimics the behaviour of $ \calC$. Starting at $ P_0 = (0,0) $. We increment (resp. decrement) the x-coordinate every time we cross an edge $ C_i \xrightarrow[]{} C_{i} + 1 $ (resp. $ C_i + 1 \xrightarrow[]{} C_{i}$). We increment (resp. decrement) the y-coordinate every time we go through an edge $ C_i \xrightarrow[]{} C_{i} + \alpha $ (resp. $ C_i + \alpha \xrightarrow[]{} C_{i}$).

When $ \calC $ is a connected cycle, $P_r = (x_r, y_r) $ the ending point of $ \gamma_\calC $ is an element of $ \LL $, that is, $ x_r + \alpha y_r \equiv 0 \, [n]$, and its Euclidean norm does not exceed $ \calC$'s length.
\vspace{0.1cm}

\begin{proposition}
     \label{prop:link_between__connected_cycle_and_L}
The coordinates of the ending  point $ P_r$ of $ \gamma_\calC $ are given by  $\left(n_1(\calC) - n_{-1}(\calC), \, n_{\alpha}(\calC) - n_{-\alpha}(\calC) \right)$ where:
   \begin{equation*}
       n_{\epsilon} \, (\calC) =  |\{\,  i \in [|0, r |] \, \,  | \,  \, C_{i+1}  - C_i \equiv \epsilon \, [n]  \, \}|  \text{ for } \epsilon \in \{ \pm \,  1 , \, \pm \, \alpha  \}
   \end{equation*}

   is the number of times we go through an edge $ C_i \xrightarrow[]{} C_i + \epsilon$ in the path $\calC$. Moreover, when $ \calC $ is a connected cycle, $ P_r \in \LL $.
\end{proposition}
\vspace{0.1cm}

\textbf{Proof of Proposition \ref{prop:link_between__connected_cycle_and_L}:}
By definition $ P_r  = \left(n_1(\calC) - n_{-1}(\calC), \, n_{\alpha}(\calC) - n_{-\alpha}(\calC) \right)$. The Euclidean norm of $ P_r$ is always smaller than $ n_1(\calC) +  n_{-1}(\calC) +  n_{\alpha}(\calC) +  n_{-\alpha}(\calC)$ which is the length of  $ \calC$. 

When $ \calC : \, C_0 \xrightarrow{}  \dots  \xrightarrow{}   C_{r}$ is a connected cycle, then $ S =  \sum_{i =  0}^{r - 1} C_{i +1} - C_i = 0 \mod n $. Moreover, \mbox{$S = (n_1(\calC) - n_{-1}(\calC)) + \alpha(n_\alpha(C) - n_{-\alpha}(\calC)) \mod n$} 
Thus, $ P_r = (n_1(\calC) - n_{-1}(\calC), n_{\alpha}(\calC) - n_{-\alpha}(\calC)) \in \LL $ \qed
\vspace{0.1cm}

Thus, proving the minimum distance bound of Theorem \ref{thm:main_theorem} is equivalent to demonstrating that connected cycles without sub-cycles have lengths greater than $ \lambda(\LL) $, the length of the shortest non-zero lattice vector.
\vspace{0.1cm}

Above, we constructed a mapping from $ \G$-walks to $ \Z^2 $-walks which ensures that any connected cycle $ \calC $ of $ \G$ is mapped to a walk $ \gamma_\calC$ of $ \Z^2$ whose ending point $ P_r$ is a lattice element of Euclidean norm greater than $ \calC$'s length. 

Next, we prove by contrapositive that if $ \calC $ is a cycle without sub-cycles that can not be expressed a sum of faces, then $ P_r$ the ending point of $ \gamma_\calC$ has a non-zero coordinate. Precisely, we will demonstrate that if, $ P_r = (0,0) $, then $ \gamma_\calC $ is a cycle that can be decomposed in a sum  of $ \Z^2 $ elementary grid squares, which in turn implies that $ \calC $ is a sum of faces of $ \G$.

\subsection{From $ \Z^2$-walks to $\G$-walks}
\label{sec: from Z^2 walks to G walks}

 While we constructed the $ \Z^2$-walk $\gamma_\calC$ to mimic the $ \G-$walk $\calC$, we can also view $\calC$ as a $ \G$-walk, starting at $C_0$, that mimics $\gamma_\calC$.
\vspace{0.1cm}

For any given $ \Z^2$-walk $  \gamma \, : \,  \gamma_0  \xrightarrow{} \dots \xrightarrow{} \gamma_{r} $ we can create a collection $  \{ (C_{\gamma} )_{t_0} \, \vert \,  t_0 \in \Z / n\Z \} $ of $ \G$-walks that replicate the sequence of steps taken by $ \gamma$. Specifically, the walk $ (C_{\gamma})_{t_0}$ is given by \mbox{$(C_{\gamma})_{t_0} : t_0  \xrightarrow{} \dots   \xrightarrow{} t_r $} where for all $i \in [|0, r - 1|]$, $ t_{i+1} \in \Z / n\Z $ is defined as follows:

\begin{itemize}
    \item $ t_{i + 1 } = t_i + 1  \mod  n $  \, \,  if $ \gamma_{i+1} - \gamma_i = (1, 0) $
\vspace{0.1cm}

    \item $ t_{i + 1 } = t_i  - 1 \mod n  $  \, \, if $  \gamma_{i+1} - \gamma_i = (- 1, 0) $
\vspace{0.1cm}

    \item  $ t_{i + 1 } = t_i + \alpha \mod n $  \, \, if $  \gamma_{i+1} - \gamma_i = ( 0, 1) $
\vspace{0.1cm}

    \item $ t_{i + 1 } = t_i - \alpha \mod n  $ \, \, if $  \gamma_{i+1} - \gamma_i = ( 0, - 1) $
    
\end{itemize}
\vspace{0.1cm}

Employing this formalism, any $\G$-walk $\calC$ can be expressed as $(\gamma_\calC)_{C_0}$, where $C_0$ is the starting point of $ \calC$. This interpretation enables us to identify conditions on $\gamma_\calC$ that guarantee that $\calC$ is a  sum of faces of $ \G$.
\vspace{0.1cm}

\begin{lemma}
\label{lemma: Z^2 cycles are sum of faces}
If \mbox{$  \gamma \, : \,   \gamma_0 = (0,0) \xrightarrow{}  \gamma_1 \xrightarrow{} \dots \xrightarrow{}  \gamma_{r} = (0,0) $} is a simple $ \Z^2$-cycle (closed loop without self-intersections), starting at the origin, then for all $  t \in \Z / n\Z, \,( C_\gamma)_t $ the associated walk in $\G$ starting at $t$ is a sum of faces of $ \G$.
\label{lemma:key_lemma}
\end{lemma}

\textbf{Proof:} We prove Lemma \ref{lemma:key_lemma} by induction.
\vspace{0.1cm}

If $ \gamma $ is a simple $\Z^2$-cycle starting at the origin that surrounds exactly 1 square of $ \Z^2 $, then $ \gamma $ is equal to that square. So, $ \forall t \in \Z / n\Z, (C_\gamma)_t  $ is one of the faces of $ \G$ starting at $t$:
\begin{itemize}
    \item (DL) : $ t \xrightarrow{} t + 1 \xrightarrow{} t + 1 + \alpha  \xrightarrow{} t + \alpha \xrightarrow{} t  $
    \vspace{0.1cm}

    \item (DR) : $ t \xrightarrow{} t + \alpha \xrightarrow{} t - 1 + \alpha  \xrightarrow{} t - 1\xrightarrow{} t  $ 
    \vspace{0.1cm}

    \item (UL) : $ t \xrightarrow{} t - \alpha  \xrightarrow{} t  + 1 - \alpha \xrightarrow{} t + 1 \xrightarrow{} t $
    \vspace{0.1cm}

    \item (UR) : $ t \xrightarrow{} t - 1 \xrightarrow{} t - 1 - \alpha  \xrightarrow{} t - \alpha \xrightarrow{} t  $
\end{itemize}
\vspace{0.1cm}

Now, assume there exists $q \geq 1$, such that for any simple $\Z^2$-loop $ \gamma$ starting at the origin and surrounding exactly \mbox{$ 1 \leq k \leq q$} $\, \Z^2$-elementary squares, all associated $\G$-walks $ (C_\gamma)_t$ are face sums.
\vspace{0.1cm}

Consider a simple $ \Z^2$-loop $ \Gamma $, going through the origin, which surrounds exactly $q + 1 $ $ \Z^2$-elementary grid squares. Let's see that for any $ t \in \Z / n\Z $, $ \Gamma_t $ is a sum of faces of $ \G$.
\vspace{0.1cm}

Let $ Int(\Gamma) $ denote the bounded connected region enclosed by $ \Gamma $.
Within $ Int(\Gamma) $, at least one of these two scenarios is true: either there is at least one grid square that shares edges with $ \Gamma$ and is not adjacent to the origin, or there are multiple such grid squares that are adjacent to the origin. We choose one of these squares, which we will call S, and let $(u,v)$ denote the coordinates of its lower-left corner.
\vspace{0.1cm}
The $ \Z^2 $-loop $ \gamma $ that surrounds the same grid squares as $ \Gamma $ but S is a simple $\Z^2$-loop going through the origin, which surrounds exactly $ q $ grid squares. Thus, $ \gamma_t $ is a sum of faces of $\G$. Since $ \Gamma_t = \gamma_t + S_{u,v} \, (t) $ where $ S_{u,v}\, (t) $ is one of the faces of $\G$ adjacent to $ t + u + \alpha v $: 
\vspace{0.1cm}

\begin{itemize}
    \item (DL) : $ t + u + \alpha v\xrightarrow{} t + u +  1 + \alpha v  \xrightarrow{} t + u + 1 + ( \alpha + 1)v \xrightarrow{} t + u + ( \alpha + 1 )v  \xrightarrow{}  t + u + \alpha v  $
    \vspace{0.1cm}

    \item (DR) : $ t + u + \alpha v \xrightarrow{} t + u + ( \alpha + 1)v + \xrightarrow{} t + u - 1 + (\alpha + 1)v  \xrightarrow{}   t + u - 1 + \alpha v     \xrightarrow{} t + u + \alpha v  $ 
    \vspace{0.1cm}
    
    \item (UR) : $ t + u + \alpha v \xrightarrow{} t + u + ( \alpha - 1)v \xrightarrow{} t  + u +  1 + (\alpha - 1)v \xrightarrow{} t + u + 1 + \alpha v  \xrightarrow{} t + u + \alpha v $
    \vspace{0.1cm}

    \item (UL) : $ t + u + \alpha v \xrightarrow{} t + u - 1 + \alpha v \xrightarrow{} t + u - 1 + (\alpha  - 1) v  \xrightarrow{}  t + u + (\alpha  - 1) v \xrightarrow{} t + u + \alpha v  $
    
\end{itemize}
\vspace{0.1cm}

Then $ \Gamma_t $ is a sum of faces of the graph. \qed
\vspace{0.1cm}

Now let's see that when $ \calC$ is a connected $ \G$-cycle without sub-cycles, if it is not the sum of faces of $ \G$, then \mbox{$ (n_1(\calC) - n_{-1} (\calC) ,\,  n_\alpha (\calC) - n_{-\alpha} (\calC)) $} is not equal to $ (0,0)$.
\vspace{0.1cm}

\begin{corollary}
     Let $ \calC \, : \,  C_0 \xrightarrow{} \dots \xrightarrow{} C_{r-1} \xrightarrow{} C_0 $ be a connected cycle in $\G$, with no sub-cycles. If $ \calC$ is not a sum of faces of $\G$ then $ n_1(\calC) - n_{-1} (\calC)\neq 0$ or $ n_\alpha (\calC) - n_{-\alpha}  (\calC) \neq 0$.
\label{corollary:corolloraly_of_key_lemma}
\end{corollary}

\textbf{Proof:} We prove the contrapositive of the corollary.

If $ \calC $ is a connected cycle of G, without sub-cycles such that $ n_1(\calC) = n_{-1} (\calC) $ and $ n_{\alpha}(\calC) = n_{-\alpha}(\calC) $, then $ \gamma_\calC $, the associated $ \Z^2$-walk, is a loop starting at $ (0,0)$. Moreover, since $\calC$ does not contain any sub-cycles, neither does $\gamma_\calC$.
Hence, by Lemma \ref{lemma: Z^2 cycles are sum of faces}, $ \calC= (\gamma_\calC)_{C_0}$ is a sum of faces of $\G$. \qed
\vspace{0.15cm}

Now, we can give a complete proof of Theorem~\ref{thm:main_theorem}.

\subsection{End of the proof of Theorem \ref{thm:main_theorem}}
\label{sec: end of main theorem}
Let $ n \geq 6 $ and $ 1 < \alpha < n - 1$ be two integers.

Let $ A , B $ and $\G$ be defined as in Section \ref{section:GB_and_graph_theory}

By Proposition \ref{prop:equivalent_GB_codes}, we may assume $ 1 < \alpha < \frac{n}{2}$. 
\vspace{0.15cm}

If $ GB(A,B,n )$ has no logical qubits, then its minimum distance is $ +\infty$ and thus Theorem \ref{thm:main_theorem} is true.
\vspace{0.1cm}

Else, we consider $ \calC $ a cycle of $\G$ that is not a sum of faces of $\G$ and whose length $ |\calC| $ is minimal for those properties. Such a cycle has to be connected and must not contain any sub-cycles. Moreover, its length $ |\calC| $ verifies:
\begin{equation*}
   |\calC| =  d(GB(a,b,n)) = n_1(\calC) + n_{-1}(\calC)  + n_{\alpha}(\calC)  + n_{-\alpha}(\calC)
\end{equation*}

where the quantities $ n_{\epsilon}(\calC) $ are defined in Proposition \ref{prop:link_between__connected_cycle_and_L}. By Corollary \ref{corollary:corolloraly_of_key_lemma} and Proposition \ref{prop:link_between__connected_cycle_and_L}, we have: 
\begin{equation*}
    P_r = ( n_1(\calC) - n_{-1}(\calC), \,  n_{\alpha}(\calC)  - n_{-\alpha}(\calC)) \in L \backslash \{ (0,0) \}
\end{equation*}

Thus $ \lambda(\LL) \leq \, ||  P_r || \leq  d(GB(A,B,n))   $ \qed 
\vspace{0.15cm}

Building on Proposition \ref{prop:equivalent_GB_codes} we can extend the bound of Theorem \ref{thm:main_theorem} to a broader class of $(2,2)$-GB codes.
\vspace{0.1cm}

\begin{corollary} 
\label{corollary: corollary of main thm}
For $ A =  1 + x^u $ and $ B =  1 + x^{v} $ two polynomials of weight two. If $ n > max(6, u, v) $ is an integer relatively prime with $ u  $ then: 
\begin{equation}
    d(GB(A,B,n)) \geq \lambda (\Z (n,0) \bigoplus \Z( -\alpha , 1)) 
\end{equation}

where $\alpha =  (  v  \mod  n)\cdot (u  \mod  n)^{-1} $
\end{corollary}
\vspace{0.1cm}

\section{Application of Theorem \ref{thm:main_theorem}}
\label{section: application of the main theorem}

Quantum coding theory seeks quantum codes with maximum minimum distance, as this improves fault tolerance. 
As all $(2,2)$-GB codes are surface codes, their minimum distance is limited to a square root scaling with the code length. Nevertheless, by applying Theorem~\ref{thm:main_theorem}, we can construct $(2,2)$-GB codes of length $2n$ and minimum distance $\Omega(n^\frac{1}{2})$ through the design of lattices $\LL = \Z(n, 0) \oplus \Z(-\alpha, 1)$ that possess a shortest non-zero vector length of $ \sqrt{n}$.

To achieve this, we use the following arithmetic result:
\vspace{0.1cm}

\begin{lemma}
\label{lemma:minoration_norme_reseau}
    Let $ 0 < \alpha  $ and let $n$ be a factor of $ 1 + \alpha^2$. 
If $ T $ is a non-zero element of $ \LL := \Z (n,0) \bigoplus \Z( -\alpha , 1)$ 
then $ n \leq || T ||^2 $. 
In particular $ n^\frac{1}{2} \leq \lambda(\LL)$
\end{lemma}
\vspace{0.1cm}

\textbf{Proof:} If $ T \neq (0,0)$ then  $ \exists (u, v) \in \Z \backslash \{(0,0) \} $ such that $ T = ( nu - \alpha v, v) $. As $ \| T \|^2 = n^2u^2 -2n \alpha u v + (1 + \alpha^2)v^2 $, $n$ divides $ \| T \|^2 $. Thus, $n \leq \| T \|^2  $ since $ \| T \|^2 > 0 $.
\qed 
\vspace{0.1cm}

Combining the results of Corollary \ref{corollary: corollary of main thm} and Proposition \ref{prop:equivalent_GB_codes}, we can conclude that when $ 1 \leq \alpha $ and $ n $ is a divisor of $1 + \alpha^2$, the $2n$-length code $GB(1 + x, \, 1 + x^{ \alpha \mod n }, \, n)$ has parameters $ [| 2n, \, 2, \, \geq \sqrt{n} |]$.
\vspace{0.1cm}

Based on the arithmetic results on the existence of a square root of $ -1$ in the cyclic group $ \Z/nZ$, these integers $n$ can be expressed as: $ n = 2^e \prod_{i = 1}^s p_i ^{\, \epsilon_i}$, where $s$ is a non-negative integer, $p_i$ are prime numbers congruent to 1 modulo 4, \mbox{$ e \in \{ 0, 1\} $} and $ \epsilon_i $ are positive integers.
\vspace{0.1cm}

Historically, the only $(2,2)$-GB codes known to exhibit a minimum distance scaling linearly with the square root of code length were the standard and optimized Kitaev codes \cite{KP13}  and those listed in Pryadko and Wang's database \cite{PW22}.

\begin{itemize}
    \item The regular Kitaev toric codes: $ [\vert 2n^2, \, 2, \, n \vert]$ given  by $GB( 1 + x $, $ 1 + x^n $, $ n^2)$ for $ n \geq 1$.

\item  Its optimized versions: $ [\vert d^2 + 1 ,\,  2, \, d  \vert]$ given by \mbox{ $GB( 1 + x^{2t^2+ 1}, \, x + x^{2t^2}, \frac{ d^2 + 1 }{2})$} where $ d = 2t + 1$ \cite{PW22}.
\end{itemize}
\vspace{0.1cm}

Our method for designing good $(2,2)$-GB codes expands the pool of such codes. To illustrate this,  we compare the number of known good $(2,2)$-GB codes of lengths less than $ 200$ prior to and following our findings:

\begin{table}[h!]
\centering
\footnotesize
\begin{tabularx}{\columnwidth}{|X|X|X|}
\hline
\makebox[\linewidth][c]{\textbf{Kitaev Codes} }& \makebox[\linewidth][c]{\textbf{\cite{PW22}} } & \makebox[\linewidth][c]{ \textbf{Our Codes} }  \\
\hline
 \makebox[\linewidth][c]{$[|  8, 2, 2 |]$} & \makebox[\linewidth][c]{--} & \makebox[\linewidth][c]{$[|  4, 2, 2 |]$} \\
\hline

\makebox[\linewidth][c]{$[|  18, 2, 3 |]$} & \makebox[\linewidth][c]{$[|  10, 2, 3 |]$} & \makebox[\linewidth][c]{$[|  10, 2, 3 |]$} \\

\hline
\makebox[\linewidth][c]{$[|  32, 2, 4 |]$} & \makebox[\linewidth][c]{$[|  22, 2, 4 |]$} & \makebox[\linewidth][c]{$[|  20, 2, 4 |]$}  \\

\hline
\makebox[\linewidth][c]{$[|  50, 2, 5 |]$} & \makebox[\linewidth][c]{$[|  26, 2, 5 |]$, $[|  38, 2, 5 |]$} & \makebox[\linewidth][c]{$[|  26, 2, 5 |]$, $[|  34, 2, 5 |]$} \\
\hline

\makebox[\linewidth][c]{$[|  72, 2, 6 |]$} & \makebox[\linewidth][c]{--} & \makebox[\linewidth][c]{$[|  52, 2, 6 |]$}  \\

\hline
\rule{0pt}{2pt} \makebox[\linewidth][c]{{$[|  98, 2, 7 |]$}} &  \rule{0pt}{2pt} \makebox[\linewidth][c]{$[|  50, 2, 7 |]$, $[|  58, 2, 7 |]$} & \makebox[\linewidth][c]{$[|  50, 2, 7 |]$}  \makebox[\linewidth][c]{$[|  58, 2, 7 |]$} \makebox[\linewidth][c]{$[|  74, 2, 7 |]$}  \\

\hline
\rule{0pt}{2pt} \makebox[\linewidth][c]{$[|  128, 2, 8 |]$} & \rule{0pt}{2pt} \makebox[\linewidth][c]{$[|  74, 2, 8 |]$} & \rule{0pt}{2pt} \makebox[\linewidth][c]{$[|  68, 2, 8 |]$, $[|  100, 2, 8 |]$} \rule[-0.25pt]{0pt}{0pt} \\
\hline

\rule{0pt}{2pt} \makebox[\linewidth][c]{$[|  162, 2, 9 |]$} & \rule{0pt}{2pt} \makebox[\linewidth][c]{$[|  82, 2, 9 |]$, $[|  106, 2, 9 |]$} & 
\makebox[\linewidth][c]{$[|  82, 2, 9 |]$} \makebox[\linewidth][c]{$[|  106, 2, 9 |]$} \makebox[\linewidth][c]{$[|  130, 2, 9 |]$}  \\
\hline

\rule{0pt}{2pt} \makebox[\linewidth][c]{$[|  200, 2, 10 |]$} & \rule{0pt}{2pt}  \makebox[\linewidth][c]{$[|  108, 2, 10 |]$} & \rule{0pt}{1pt}  \makebox[\linewidth][c]{$[|  116, 2, 10 |]$} \makebox[\linewidth][c]{$[|  122, 2, 10 |]$} \rule[-0.25pt]{0pt}{0pt}\\

\hline
\rule{0pt}{2pt} \makebox[\linewidth][c]{--} & \rule{0pt}{1pt} \makebox[\linewidth][c]{ $[|  122, 2, 11 |]$} \makebox[\linewidth][c]{$[|  134, 2, 11 |]$ } & \rule{0pt}{2pt} \makebox[\linewidth][c]{$[|  146, 2, 11 |]$} \rule[-0.25pt]{0pt}{0pt} \\
\hline

\rule{0pt}{2pt} \makebox[\linewidth][c]{--} & \rule{0pt}{2pt} \makebox[\linewidth][c]{$[|  166, 2, 12 |]$}  & \rule{0pt}{2pt} \makebox[\linewidth][c]{$[|  148, 2, 12 |]$} \rule[-0.25pt]{0pt}{0pt} \\
\hline

\rule{0pt}{2pt} \makebox[\linewidth][c]{--} & \rule{0pt}{2pt} \makebox[\linewidth][c]{$[|  170, 2, 13 |]$} & \makebox[\linewidth][c]{$[|  170, 2, 13 |]$} \makebox[\linewidth][c]{$[|  178, 2, 13 |]$} \makebox[\linewidth][c]{$[|  194, 2, 13 |]$}  \\
\hline
\end{tabularx}
\label{tab:code-comparison}
\end{table}

\section*{Acknowledgements}
The authors wish to thank Gilles Zémor for insightful discussions on this work.


\end{document}